\documentclass[aps,amssymb,pra,showpacs,twocolumn]{revtex4}

\usepackage{epsf}
\usepackage{amsmath}
\usepackage{graphicx}
                                                                                                                          
\begin{document}                                                       
                                                     
\title{Euclidean resonance in a magnetic field}

\author{B. Ivlev} 

\affiliation
{Department of Physics and Astronomy and NanoCenter\\
University of South Carolina, Columbia, SC 29208\\
and\\
Instituto de F\'{\i}sica, Universidad Aut\'onoma de San Luis Potos\'{\i}\\
San Luis Potos\'{\i}, S. L. P. 78000 Mexico}

\begin{abstract}

An analogy between Wigner resonant tunneling and tunneling across a static potential barrier in a static magnetic field is
found. Whereas in the process of Wigner tunneling an electron encounters a classically allowed region, where a discrete 
energy level coincides with its energy, in the magnetic field a potential barrier is a constant in the direction of 
tunneling. Along the tunneling path the certain regions are formed, where, in the classical language, the kinetic energy of
the motion perpendicular to tunneling is negative. These regions play a role of potential wells, where a discrete energy 
level can coincide with the electron energy. Such phenomenon, which occurs at the certain magnetic field, is called 
Euclidean resonance and substantially depends on a shape of potential forces in the direction perpendicular to tunneling. 
Under conditions of Euclidean resonance a long distance underbarrier motion is possible. 

\end{abstract} \vskip 1.0cm
   
 
\maketitle

\section{INTRODUCTION}
\label{introduction}
It is known that a long distance motion under a classical static potential barrier is impossible excepting a short WKB 
(Wentzel, Kramers, and Brillouin) penetration \cite{LANDAU1}. This is true in a one-dimensional case. In  two dimensions a
situation can be more complicated since, besides a kinetic energy in the direction of tunneling, there is also a 
transverse kinetic energy (in the perpendicular direction). In terms of Schr\"{o}dinger equation, those parts are 
proportional to second derivatives of a wave function. An underbarrier propagating motion, in principle, is possible if 
the transverse kinetic energy would be negative in order to compensate the total negative energy allowing a positive 
kinetic energy in the direction of tunneling. 

Let us specify the problem in more details. Suppose that the tunneling path is in the $x$-direction and away from ends of 
the path the total potential energy is $u(y)$. This function is even and has the minimum $u(0)=0$. The magnetic field is 
directed along the $z$-axis. So the negative electron energy $E$ corresponds to an underbarrier motion. One can write down
the effective Schr\"{o}dinger equation for a motion in the $x$-direction for the wave function $\varphi(x)=\psi(x,0)$
\begin{equation} 
\label{201}
-\frac{\hbar^{2}}{2m}\frac{\partial^{2}\varphi(x)}{\partial x^{2}}+U(x)\varphi =E\varphi,
\end{equation}
where $U(x)=-(\hbar^{2}/2m\psi)\partial^{2}\psi/\partial y^{2}$ is taken at $y=0$. We use the gauge $\vec A=\{-Hy,0,0\}$. 
The wave function can be expressed through the modulus and the phase as $\psi =|\psi|\exp(i\chi)$. Since the potential 
$u(y)$ is symmetric, the modulus of the wave function is even and the phase $\chi$ is odd with respect to $y$. For this 
reason, the effective potential 
\begin{equation} 
\label{204}
U(x)=\frac{\hbar^{2}}{2m}\left[\left(\frac{\partial\chi}{\partial y}\right)^{2}
-\frac{1}{|\psi|}\frac{\partial^{2}|\psi|}{\partial y^{2}}\right]_{y=0}
\end{equation}
is real. 
          
In the absence of the magnetic field there is nothing surprising. An electron is localized at the center of the potential 
$u(y)$, at $y=0$, where $|\psi|$ has a maximum. Therefore, the both terms in Eq.~(\ref{204}) are positive. In this case 
$U(x)$ is positive and the underbarrier motion with total negative energy $E$ relates to an exponential decay of the wave 
function of the WKB type. 

In the magnetic field the situation can be substantially different. Despite the minimum of the potential energy $u(y)$ is 
at $y=0$, due to Lorentz forces, maxima of the electron distribution can be shifted symmetrically apart from the line 
$y=0$ (disjoining effect). In this case $\partial^{2}|\psi|/\partial y^{2}$ is positive at $y=0$ and the second term in 
Eq.~(\ref{204}) is negative. Such disjoining electron distribution was pointed out in Refs.~\cite{SHKL3} and 
\cite{IVLEV7}. We discuss the disjoining effect below.

It is hard to conclude in advance that the disjoining distribution of the electron density away form the center, $y=0$, 
(the second term in Eq.~(\ref{204})) is sufficient to drive the total $U(x)$ to a negative value. Nevertheless, on the 
basis of the analytical solution of Ref.~\cite{IVLEV7}, it is found that the potential $U(x)$ has a form of negative 
potential wells. One can compare positions of discrete energy levels in those wells with the value of the electron energy 
$E<0$ according to Eq.~(\ref{201}). If, under variation of the magnetic field, some level in the well $U(x)$ approaches 
the energy $E$ this should result in the resonance phenomenon called Euclidean resonance \cite{IVLEV7}. 

Euclidean resonance in a magnetic field reminds the phenomenon of Wigner resonant tunneling \cite{LANDAU1}, when in the
middle a potential barrier there is a well with a level close to the particle energy. But an essential difference is that 
in Euclidean resonance the initial system is homogeneous along the direction of tunneling and the effective wells are 
formed by an intrinsic mechanism due to the transverse motion. 

When a negative underbarrier energy $E$ is fixed, one can compare it with the energy parameter $m\omega^{2}_{c}a^{2}$, 
where the cyclotron frequency is $\omega_{c}=|e|H/mc$ and $a$ is a typical spatial scale of the potential $u(y)$. In the
limit of high magnetic fields, $m\omega^{2}_{c}a^{2}\gg |E|$, energy levels in the potential well $U(x)$ are of the order 
of $-m\omega^{2}_{c}a^{2}$ which substantially lower then the energy $E$. In this case there is no levels coincidence. 

Upon the reduction of the magnetic field, the two energies become of the same order of magnitude, 
$-m\omega^{2}_{c}a^{2}\sim E$, which indicates a principal possibility of levels coincidence. As calculations show, this
occurs at the magnetic field $H_{R}$ which can be called resonance magnetic field. An analytical form of the potential 
$u(y)$ in the plane of complex $y$ plays a crucial role. For example, Euclidean resonance is absent for a pure quadratic 
$u(y)$. This sensitivity is a consequence of interference of underbarrier cyclotron paths which are reflected from the 
potential $u(y)$. 

Near the resonance field $H_{R}$ the underbarrier exponential decay of the wave function becomes weak due to the resonant 
connection of different potential wells. At $H=H_{R}$ there is no an exponential decay (perhaps, a power law). This means 
that the particle can penetrate over a long distance under the classical barrier. We emphasize that the particle energy is
strictly below the potential barrier at each point of it. Euclidean resonance (formation of the long range coherence) 
constitutes a phenomenon which can be considered as an opposite pole with respect to Anderson localization (destruction of
the long range coherence) \cite{SEVA}.  

In this paper we analytically calculated the coordinate dependence of the wave function $\psi(x,y)$ and established the 
Euclidean resonance condition using classical complex trajectories. We proposed an interpretation of the phenomenon. A 
topological vortex state under the barrier is formed which results in the effective potential well. Euclidean resonance 
corresponds to coincidence of a level in that well with the electron energy. 
           
Tunneling in a magnetic field was addressed in 
Refs.~\cite{SHKL1,IVLEV7,VOLOVIK,SHKL2,SHKL3,THOUL,SHKL4,RAIKH1,RAIKH2,GESH,BLATT,GOROKH}.

\section{FORMULATION OF THE PROBLEM}
\label{formulation}
We consider an eigenstate with a negative energy $E_{1}$ of the Schr\"{o}dinger equation 
\begin{equation} 
\label{1}
-\frac{\hbar^{2}}{2m}\left(\frac{\partial}{\partial x}-\frac{iy}{l^{2}}\right)^{2}\psi-
\frac{\hbar^{2}}{2m}\frac{\partial^{2}\psi}{\partial y^{2}}+\left[V(x)+u(y)\right]\psi =E_{1}\psi
\end{equation}
in the magnetic field directed along the $z$-axis. The magnetic length is introduced \cite{LANDAU1}
\begin{equation}
\label{2}
l=\sqrt{\frac{\hbar}{m\omega_{c}}}.
\end{equation}
The $x$-part of the potential is the negative $\delta$-well, $V(x)=-\hbar\sqrt{2|E|/m}\delta(x)$. The $y$-part has the form
\begin{equation}
\label{4a}
u(y)=u_{0}\left(\frac{y}{a}\right)^{4N},
\end{equation}
where $N$ is a large integer number. The potential (\ref{4a}) represents approximately two infinite potential walls at the
points $y=\pm a$. In the absence of the magnetic field the lower discrete energy level in the potential $u(y)$ can be 
estimated as $\hbar^{2}/ma^{2}$. The energies $|E|$ and $\hbar\omega_{c}$ are supposed to be large compared to that energy
\begin{equation} 
\label{0}
\frac{\hbar^{2}}{ma^{2}}\ll |E|,\hspace{1cm}l\ll a.
\end{equation}
The second condition (\ref{0}) can also be expressed in the form $1\ll n$, where
\begin{equation} 
\label{0a}
n=\frac{Ha^{2}}{\Phi_{0}}
\end{equation}
is a number of flux quanta $\Phi_{0}=\pi c\hbar/e$ of the total magnetic flux through the area $a^{2}$. 

The boundary condition for the region $x>0$ has the form
\begin{equation} 
\label{4}
\hbar\left(\frac{\partial}{\partial x}-\frac{iy}{l^{2}}\right)\psi(x,y)\bigg |_{x=0}=-\sqrt{2m|E|}\psi(0,y).
\end{equation} 
Under the semiclassical conditions (\ref{0}) the eigenvalue $E_{1}$ almost coincides with $E$.

\section{THE HAMILTON-JACOBI EQUATION}
One can always specify the wave function in the form
\begin{equation} 
\label{5}
\psi(x,y)=\exp\left(\frac{iS(x,y)}{\hbar}\right),
\end{equation}
where $S(x,y)$ satisfies the equation
\begin{equation} 
\label{6}
\frac{1}{2m}\left(\frac{\partial S}{\partial x}-m\omega_{c}y\right)^{2}+
\frac{1}{2m}\left(\frac{\partial S}{\partial y}\right)^{2}-\frac{i\hbar}{2m}\nabla^{2}S=E.
\end{equation}
The boundary condition (\ref{4}) is now transformed into 
\begin{equation} 
\label{7}
\frac{\partial S(x,y)}{\partial x}\bigg|_{x=0}=i\sqrt{2m|E|}+m\omega_{c}y.
\end{equation}
Below we measure $x$ and $y$ in the units of the cyclotron length
\begin{equation} 
\label{8}
L=\sqrt{\frac{2|E|}{m\omega^{2}_{c}}}=l\sqrt{\frac{2|E|}{\hbar\omega_{c}}}.
\end{equation}

The new function $\sigma(x,y)$ is introduced
\begin{equation} 
\label{9}
S(x,y)=i\frac{2|E|}{\omega_{c}}\sigma(x,y)
\end{equation}
which obeys the equation
\begin{equation} 
\label{10}
\left(\frac{\partial\sigma}{\partial x}+iy\right)^{2}+\left(\frac{\partial\sigma}{\partial y}\right)^{2}
-\frac{\hbar\omega_{c}}{2|E|}\left(\frac{\partial^{2}\sigma}{\partial x^{2}}
+\frac{\partial^{2}\sigma}{\partial y^{2}}\right)=1
\end{equation}
with the boundary condition
\begin{equation} 
\label{11}
\frac{\partial\sigma(x,y)}{\partial x}\bigg|_{x=0}=1-iy.
\end{equation}
The reflection occurs at the points $y=\pm\alpha$, where
\begin{equation} 
\label{11a}
\alpha=\sqrt{\frac{m\omega^{2}_{c}a^{2}}{2|E|}}=\frac{a}{L}.
\end{equation}
We consider below the limit of a relatively small magnetic field $\hbar\omega_{c}\ll |E|$. In this case one can write 
Eq.~(\ref{10}) in the form of Hamilton-Jacobi equation 
\begin{equation} 
\label{11b}
\left(\frac{\partial\sigma}{\partial x}+iy\right)^{2}+\left(\frac{\partial\sigma}{\partial y}\right)^{2}=1,
\end{equation}
where $\sigma$ plays a role of a classical action. 

\section{SOLUTION OF THE HAMILTON-JACOBI EQUATION}
\label{solution}
\begin{figure}
\includegraphics[width=8cm]{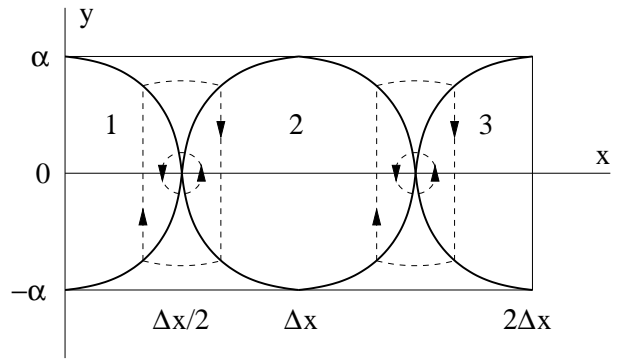}
\caption{\label{fig1}The reflection-induced curves restrict the regions (1), (2), and (3), where the reflectionless 
approximation holds. The dashed paths show the electric current of vortices.}
\end{figure}

The equation (\ref{11b}) can be solved by the variation of constants as described in \cite{LANDAU2}. The general integral
of the Hamilton-Jacobi equation has the form
\begin{equation} 
\label{12}
\sigma(x,y)=vx+\int^{y}_{0}dy_{1}\sqrt{1+(y_{1}-iv)^{2}}+F(v),
\end{equation}
where the certain function $v(x,y)$ is introduced which should be determined from the condition (independence of $\sigma$
on $v$)
\begin{equation} 
\label{13}
x-i\int^{y}_{0}\frac{(y_{1}-iv)dy_{1}}{\sqrt{1+(y_{1}-iv)^{2}}}+\frac{\partial F(v)}{\partial v}=0.
\end{equation}
According to that, the following relations hold
\begin{equation} 
\label{14}
\frac{\partial\sigma(x,y)}{\partial x}=v(x,y),\hspace{0.3cm}\frac{\partial\sigma(x,y)}{\partial y}=
\sqrt{1+\left[y-iv(x,y)\right]^{2}}.
\end{equation}
The function $F(v)$ should be determined from the condition (\ref{11}) which now reads 
\begin{equation} 
\label{15}
v(0,y)=1-iy.
\end{equation}
If we express $y$ through $v(0,y)$ and insert into Eq.~(\ref{13}) at $x=0$, we obtain the functional dependence 
\begin{equation} 
\label{16}
\frac{\partial F(v)}{\partial v}=\sqrt{v^{2}-1}.
\end{equation}
Eq.~(\ref{13}), which determines the function $v(x,y)$, takes the form
\begin{equation} 
\label{17}
x=i\int^{y}_{iv-i}\frac{(y_{1}-iv)dy_{1}}{\sqrt{1+(y_{1}-iv)^{2}}}.
\end{equation}
As follows from Eq.~(\ref{17}), $v(x,y)=\sqrt{1+x^{2}}-iy$. With this definition, Eq.~(\ref{14}) yields
\begin{equation} 
\label{18}
\frac{\partial\sigma(x,y)}{\partial x}=\sqrt{1+x^{2}}-iy,\hspace{0.3cm}\frac{\partial\sigma(x,y)}{\partial y}=ix.
\end{equation}
The form (\ref{12}) and Eqs.~(\ref{16}) - (\ref{18}) hold at the certain part of the $\{x,y\}$-plane. The integration 
variable $y_{1}$ in Eq.~(\ref{17}) varies between the limits $y$ and $i(\sqrt{1+x^{2}}-1)+y$. Our reflectionless 
approach ($u(y)\simeq 0$) is valid when $|y_{1}|<\alpha$. This condition reads
\begin{equation} 
\label{19}
\left(\sqrt{1+x^{2}}-1\right)^{2}+y^{2}<\alpha^{2}.
\end{equation}
The condition (\ref{19}) defines the area in $\{x,y\}$-plane, marked as (1) in Fig.~\ref{fig1}. The area (1) is restricted
by the solid curve which can be treated as reflection-induced one. This means that it accumulates interfering waves 
reflected from the walls.
\begin{figure}
\includegraphics[width=7cm]{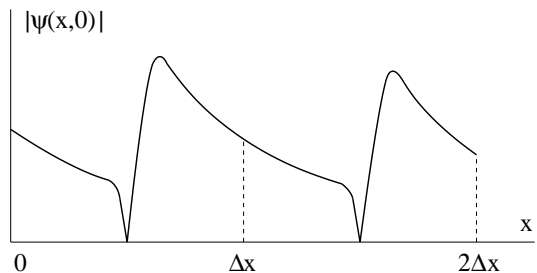}
\caption{\label{fig2}The modulus of the wave function decays slightly, when $H$ is close to $H_{R}$. The long range 
semiclassical parts are connected by the short vortex regions, of the width of $l^{2}/a$, localized near the points 
$x=\Delta x/2$ and $3\Delta x/2$, where $\psi=0$.}
\end{figure}
The period of the structure is $\Delta x=2\sqrt{\alpha(2+\alpha)}$. In the physical units
\begin{equation} 
\label{22}
\Delta x=2a\left(1+\sqrt{\frac{8|E|}{m\omega^{2}_{c}a^{2}}}\right)^{1/2}.
\end{equation}
According to Eqs.~(\ref{5}), (\ref{9}), and (\ref{18}), at $0<x<\Delta x/2$, in the physical units, 
\begin{equation} 
\label{19a}
|\psi(x,0)|\sim\exp\left[-\frac{2|E|}{\hbar\omega_{c}}\ln\left(\frac{x}{L}+\sqrt{1+\frac{x^{2}}{L^{2}}}\right)\right]. 
\end{equation}

On the distance $\delta$ around the point $x=\Delta x/2$, $y=0$ the semiclassical approximation breaks down. The scale 
$\delta$ can be estimated from Eq.~(\ref{10}). The two terms, $p^{2}_{x}$ and 
$(\hbar\omega/|E|)\partial p_{x}/\partial x$, where $p_{x}=\partial\sigma/\partial x$, should be of the same order of 
magnitude within the non-semiclassical region. In physical units this gives the estimate $\delta\sim l^{2}/a$. There is a
clear interpretation of the quantum length $\delta$. The cyclotron length (\ref{8}) is estimated as $L\sim p/m\omega_{c}$.
The length $\delta$ can be obtained from the same relation if to put $L\sim a$ and the quantum uncertainty condition 
$p\sim\hbar/\delta$. The length $\delta$ has a meaning of a quantum uncertainty in position of a center of a cyclotron 
orbit (Lorentz drift uncertainty). 

How to go beyond the region (1) in Fig.~\ref{fig1}? 

In the semiclassical problem of a one-dimensional overbarrier motion there is an effect of formation of reflected wave at
the certain point $x$ (Stokes phenomenon). At this point the Stokes line, which starts at the complex point $x_{R}$, 
intersects the real axis \cite{HEADING}. The easiest way to obtain the reflected wave is to go in the complex plane up to 
$x_{R}$, where two branches (incident and reflected) merge, and to return to the real axis. This method accounts for a 
delicate interference of partial de Broglie waves emitted by the particle.

In our problem the situation is qualitatively similar. The two branches, related to the regions (1) and (2) in 
Fig.~\ref{fig1}, also merge at the certain complex point $x=\Delta x/2,~y=-i\eta$. At that point the root in 
Eq.~(\ref{14}) turns to zero if to account for the entire potential (\ref{4a}). The solution, generated in this way in the
region (2), is $\sigma_{2}(x,y)=\sigma(x-\Delta x,y)+{\rm const}$. The constant can be determined by the same method of
going in the complex plane of the variables. This program is performed in the next Section. 

\section{CLASSICAL TRAJECTORIES}
The classical path from the region (1) to the region (2) in Fig.~\ref{fig1} goes through the complex plane of coordinates 
between the physical points $\{x=0,\hspace{0.05cm}y=0\}$ and $\{x=\Delta x,\hspace{0.05cm}y=0\}$. Along this path, $x$ is 
real but $y=-i\eta$ is imaginary. This path corresponds to a trajectory in imaginary time. The proper formalism is 
developed in Ref.~\cite{IVLEV7} and results in
\begin{equation} 
\label{20}
\bigg|\frac{\psi(\Delta x,0)}{\psi(0,0)}\bigg|^{2}\sim\exp(-A),
\end{equation}
where 
\begin{equation} 
\label{20a}
A=A_{WKB}(\Delta x)-\frac{2m}{\hbar}\int d\tau\left(\frac{\partial\eta}{\partial\tau}\right)^{2}.
\end{equation}
The integral in Eq.~(\ref{20a}) is taken along one period of the periodic motion described by the classical equation
\begin{equation} 
\label{20b}
\frac{m}{2}\left(\frac{\partial\eta}{\partial\tau}\right)^{2}-\frac{m\omega^{2}_{c}}{2}\left(\eta+L\right)^{2}
+u(-i\eta)=E.
\end{equation}
The last term in Eq.~(\ref{20a}) reduces the total action and can be interpreted as one originated from the transverse 
kinetic energy as it is discussed in Sec.~\ref{introduction}. The action (\ref{20a}) has the form
\begin{equation} 
\label{21}
A=\left(1-\int^{\alpha}_{0}d\eta\sqrt{\frac{\eta(2+\eta)}{\alpha(2+\alpha)}}\right)A_{WKB}(\Delta x),
\end{equation}
where the WKB action is $A_{WKB}(\Delta x)=2\Delta x\sqrt{2m|E|}/\hbar$. The wave function in the region (1) in 
Fig.~\ref{fig1} is given by Eqs.~(\ref{5}), (\ref{9}), (\ref{12}), (\ref{16}), and (\ref{17}). In the region (2) the 
action $\sigma_{2}(x,y)$ is expressed through the function $\sigma(x,y)$ by the equation
\begin{equation} 
\label{23}
\sigma_{2}(x,y)-\sigma(x-\Delta x,y)=\Delta x-2\int^{\alpha}_{0}d\eta\sqrt{(\eta +1)^{2}-1}.
\end{equation}
The same relation holds for the region (3) in Fig.~\ref{fig1}, namely, the difference 
$\sigma_{3}(x,y)-\sigma_{2}(x-\Delta x,y)$ coincides with the right-hand side of Eq.~(\ref{23}). 

The function $|\psi(x,y)|$ does not depend on $y$ in the regions (1), (2), and (3). The modulus, $|\psi(x,0)|$, is plotted 
in Fig.~\ref{fig2}. The vortex cores are located at the points $x=\Delta x/2$ and $x=3\Delta x/2$, where $\psi=0$ 
\cite{LANDAU1}. This is analogous to Fig.~\ref{fig3}.

Under the increase of the magnetic field the parameter $\alpha$ increases and, when it reaches the value 
$\alpha_{R}\simeq 1.66$, the right-hand sides of equations (\ref{21}) and (\ref{23}) formally turn to zero. This occurs at 
the resonance magnetic field
\begin{equation} 
\label{24}
H_{R}=\frac{c\sqrt{2m|E|}}{|e|a}\alpha_{R}.
\end{equation}
One can simply show, that close to $H_{R}$
\begin{equation} 
\label{25}
\bigg|\frac{\psi(R,0)}{\psi(0,0)}\bigg|^{2}\sim\exp\left(-0.94\frac{H_{R}-H}{H_{R}}A_{WKB}(R)\right),
\end{equation}
where $R$ is an integer number of $\Delta x$. Under the resonance condition, $H=H_{R}$, the spatial scale in 
Fig.~\ref{fig1} and Fig.~\ref{fig2} is $\Delta x\simeq 2.97a$. Our semiclassical method is applicable when the action is 
formally large. So there is a restriction $1/A_{WKB}\ll (H_{R}-H)/H_{R}\ll 1$. To really approach the resonance field 
$H_{R}$ one should go beyond the semiclassical approximation.  

\section{ELECTRIC CURRENT UNDER THE BARRIER}
\label{current}
The electric current $\vec j=-e^{2}|\psi|^{2}{\vec Q}/mc$ is expressed through the gauge invariant vector potential 
$\vec Q=\vec A+Hl^{2}\nabla\chi$ which depends on the phase $\chi=-2|E|{\rm Im}\sigma/\hbar\omega_{c}$ of the wave 
function $\psi=|\psi|\exp(i\chi)$ \cite{LANDAU1}. With the gauge $\vec A=\{-Hy,0,0\}$ used, there is only $y$-component of 
$\vec Q$ in the regions (1), $-Hx$, and in the region (2), $-H(x-\Delta x)$. The electric current distribution is shown by
the dashed curves in Fig.~\ref{fig1}. 

\begin{figure}
\includegraphics[width=8cm]{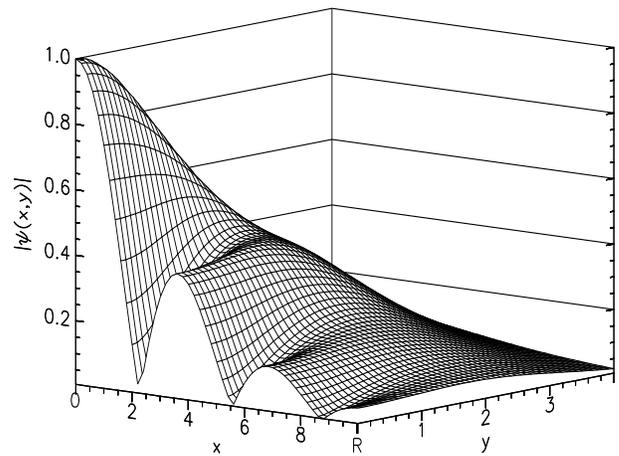}
\caption{\label{fig3}The modulus of the wave function from Ref.~\cite{IVLEV7}, when the magnetic field is high, 
$m\omega^{2}_{c}a^{2}\gg |E|$. The plot is symmetric with respect to the line $y=0$. The variables $x$ and $y$ are 
measured in the units of $2l^{2}/a$.}
\end{figure}

There are topological vortices at the points $\{\Delta x/2,0\}$ and $\{3\Delta x/2,0\}$, as in Fig.~\ref{fig3}. The 
vorticity along the dashed small loop, of the size $\delta$, in Fig.~\ref{fig1} 
\begin{equation} 
\label{25a}
\oint{\vec Q}d{\vec l}=-\Phi_{0}
\end{equation}
is determined by the topological properties \cite{DEGENNES}. The vorticity along the large dashed loop, of the size $a$, 
changes sign in order to provide the total positive magnetic flux
\begin{equation}
\label{25b}
\Phi=\Phi_{0}+\oint{\vec Q}d{\vec l}
\end{equation}
through the area restricted by the large loop. In Eqs.~(\ref{25a}) and (\ref{25b}) the contours of integration are
counter-clockwise. In contrast to vortices in superconductors, in this case there is no Meissner screening which would 
cancel the second term in Eq.~(\ref{25b}) at a large distance resulting in the total flux $\Phi_{0}$ per one vortex 
\cite{DEGENNES}. 

One should emphasize opposite roles of different parts of the vortices. At the outer part of the vortex (the large dashed 
loop in Fig.~\ref{fig1}) the Lorentz force is directed inside the loop leading to joining of the electron density to the 
vicinity of the vortex. This results in the peaks of $|\psi|$ in Fig.~\ref{fig2}. At the inner part (the small dashed loop)
the Lorentz force is directed outside the loop leading to the local disjoining of the electron density in the close 
vicinity to the vortex topological point ($\psi=0$) in Fig.~\ref{fig2}. A contribution of the inner part to the total
angular momentum is small. 

The current paths are continued outside the regions (1), (2), and (3) in Fig.~\ref{fig1}, where we do not know a detailed 
form of the wave function. The vortex structure of the wave function is a consequence of a specific analytical form of the
potential $u(y)$ \cite{IVLEV7} and does not depend on the magnetic field. For example, for a quadratic $u(y)$ topological
vortices under the barrier are absent and there are smooth current curves only. 

\section{CHOICE OF $u(y)$}
We use the potential $u(y)$ determined by Eq.~(\ref{4a}) with a large $N$ which allows to treat the potential (\ref{4a}) 
as the infinite walls at $y=\pm a$ and $y=\pm ia$. For the semiclassical approach $N$ should not be too large. The typical
scale near the wall, $\delta y\sim a/N$, has to be not very short satisfying the semiclassical condition (\ref{0}) with 
$\delta y$ instead of $a$. This leads to the condition $N\ll n$, where the number of magnetic flux quanta $n$ is 
determined by Eq.~(\ref{0a}). That condition does not contradict to $1\ll N$ since the number $n$ is large. 

The potential walls should not be infinitely steep. Otherwise, in the limit of $N\rightarrow\infty$ (perfectly rectangular 
potential barriers) the semiclassical approach is not valid. In the perfectly rectangular limit, the condition 
$|y_{1}|<\alpha$, resulting in the Eq.~(\ref{19}), is substituted by ${\rm Re}\hspace{0.05cm}y_{1}<\alpha$ which yields
the condition $y^{2}<\alpha^{2}$. The consequence is that the reflection-induced curves in Fig.~\ref{fig1} are degenerated
into two lines $y=\pm \alpha$ and the underbarrier vortex state is not formed. The effect exists for any $N=1,2,...$ in 
formula (\ref{4a}) (see also the discussion of a form of $u(y)$ in Ref.~\cite{IVLEV7}). We use a large $N$ since it 
simplifies solution of the Hamilton-Jacobi equation allowing the reflectionless approach at some parts of the 
$\{x,y\}$-plane.

\section{INTERPRETATION}
\label{general}
The maxima of the electron density are shifted apart from the line $y=0$ due to the disjoining effect near the vortex 
cores. An example of such distribution is shown in Fig.~\ref{fig3}. The path $y=0$ provides a convenient indication of the
resonance since along this path the effective potential (\ref{204}) is real which allows the interpretation in terms of 
the conventional Schr\"{o}dinger equation (\ref{201}). 

The potential $U(x)$ can be calculate analytically in the case of the high magnetic field, $m\omega^{2}_{c}a^{2}\gg |E|$, 
considered in Ref.~\cite{IVLEV7}, where it was taken $u(y)=u_{0}\left(y^{2}/a^{2}+y^{4}/a^{4}\right)$. The decaying wave 
function is illustrated in Fig.~\ref{fig3}. According to Eq.~(\ref{201}), apart from the vortex singularity, $x\neq 0$, 
the potential $U(x)$ can be written in the form
\begin{equation} 
\label{204a}
U(x)=\frac{\hbar^{2}}{2m|\psi(x,0)|}\frac{\partial^{2}|\psi(x,0)|}{\partial x^{2}}+E.
\end{equation}
As follows from Fig.~\ref{fig3}, at the extrema of $|\psi(x,0)|$ the function (\ref{204a}) is negative. In other words, 
the second term in Eq.~(\ref{204}) dominates. The derivative $\partial^{2}|\psi|/\partial y^{2}$ at $y=0$ is positive
since the electron density is disjoined near the line $y=0$. 
           
It is easy to calculate $U(x)$ on the basis of the results of Ref.~\cite{IVLEV7}. Inside the barrier this potential has the
form
\begin{equation} 
\label{205}
U(x)=\frac{m\omega^{2}_{c}a^{2}}{4}\left[1+\sqrt{3}
\tan\left(\frac{a(x-x_{0})}{2l^{2}}\right)\right].
\end{equation}
The shift $x_{0}$ is chosen in order to get the singularities in the expression (\ref{205}) at the vortex positions. The 
\begin{figure}
\includegraphics[width=7.5cm]{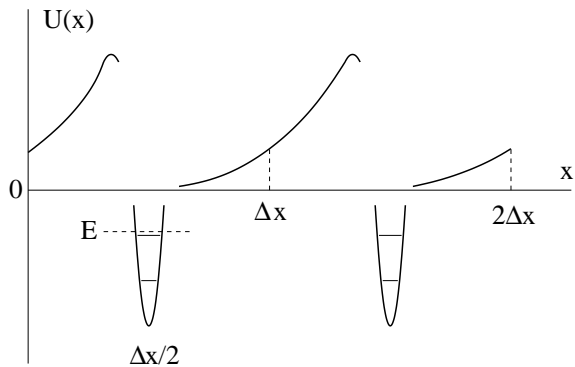}
\caption{\label{fig4}The potential $U(x)$ close to the condition of Euclidean resonance. The parabolic semiclassical 
segments are connected at vicinities of the points $\Delta x/2$ and $\Delta 3x/2$ by the non-semiclassical wells of the 
width $\delta\sim a/n\ll\Delta x\sim a$ which are shown schematically.}
\end{figure}
potential $U(x)$ contains discrete energy levels of the order of $-m\omega^{2}_{c}a^{2}$ which are smeared out to the 
bands of the same order of magnitude. Since $E$ is much smaller than that scale, it is not placed inside an allowed energy
band. For this reason, the state is exponentially decayed. 

The distribution of $|\psi(x,y)|$ in Fig.~\ref{fig3} corresponds to the limit $m\omega^{2}_{c}a^{2}\gg |E|$. As usually, 
it should be qualitatively similar at the border of applicability, $m\omega^{2}_{c}a^{2}\sim |E|$. In this limit, the 
topological vortex structure has to be of the same type as in Fig.~\ref{fig3} and the energy levels in the well 
$-m\omega^{2}_{c}a^{2}$ are of the order of $E$. In this case an energy coincidence is expected which results in the 
resonance.

At the magnetic field $m\omega^{2}_{c}a^{2}\sim |E|$ the periodic potential $U(x)$ (\ref{204}) can be evaluated from the 
results of Sec.~\ref{solution}. At $\Delta x/2<x<3\Delta x/2$ the potential has the form 
$U(x)=m\omega^{2}_{c}(x-\Delta x/2)^{2}/2$ shown in Fig.~\ref{fig4}. These periodic parabolic segments are connected by
potential wells located close to the points $\Delta x/2$ and $3\Delta x/2$ within the non-semiclassical interval 
$\delta\sim l^{2}/a$ estimated in Sec.~\ref{solution}. The total $U(x)\sim E$ is plotted schematically in Fig.~\ref{fig4}.
The discrete energy levels, shown in Fig.~\ref{fig4}, are of the order of $-m\omega^{2}_{c}a^{2}$ and slightly smeared out 
into narrow energy bands. Now one can expect a coincidence of the energy level $-m\omega^{2}_{c}a^{2}$ with the electron
energy $-E$ which is of the same order of magnitude. As we know from the calculations, it happens at $H=H_{R}$ (Euclidean 
resonance).

\section{DISCUSSIONS}
We established the analogy between Euclidean resonance and conventional resonant Wigner tunneling by means of the 
effective potential $U(x)$, where a level coincides with the negative electron energy. It is clear that $U(x)$ should be 
negative at least at some regions. The necessary condition for this property, as follows from Eq.~(\ref{204}), is 
positivity of $\partial^{2}|\psi|/\partial y^{2}$ at some regions of $x$ on the line $y=0$. In other words, the electron
density should be disjoined at those regions. The disjoining effect is provided by the Lorentz force at the inner part of
topological vortices, as noted in Sec.~\ref{current}. So the logical chain for explanation of Euclidean resonance is the 
following:
\begin{itemize}
\item creation of topological vortices
\item disjoined electron density due to the Lorentz force produced by the vortex current
\item a local positive $\partial^{2}|\psi|/\partial y^{2}$ as a result of the disjoined distribution
\item an effective potential $U(x)$ with local negative potential wells
\item coincidence of a level in the well with the electron energy (Euclidean resonance as a form of resonant Wigner 
tunneling)
\end{itemize}
In this sequence a transition from one item to another is logical. The last item provides the analogy we explore. The only 
unclear item is the initial one, creation of topological vortices. Actually, this question is {\it heart of the problem}. 

Topological vortices under the barrier are created, for example, when $u(y)\sim y^{4}$ and are not created in the case of 
a quadratic $u(y)$. There is the mathematical way to understand this (analytical properties of the function $u(y)$ in the 
complex plane) but it is impossible to propose general physical arguments to explain that difference. Indeed, why for some
potential $u(y)$ the underbarrier state has a different topology? As a rule, physical arguments do not work for explanation
of complicated interference. An example is non-reflectivity of certain potentials due to interference of emitted partial 
de Broglie waves \cite{LANDAU1}. 

Euclidean resonance provides an example in condensed matter physics when the result is unexpected and cannot be predicted 
prior to calculations. Also a dependence of topology on form of the potential cannot be explained by hand waving 
arguments. 

Euclidean resonance means a formation of the long range coherence among effective potential wells which is opposite to 
Anderson localization when the coherence is destroyed \cite{SEVA}. One should note that Euclidean resonance is not only a 
property of a static potential barrier in a magnetic field but also occurs in tunneling through nonstationary barriers 
\cite{IVLEV2,IVLEV3,IVLEV4,IVLEV5,IVLEV7}. In the both cases an important issue is a formation of a phase of the wave 
function in the process of the underbarrier motion. This leads to an interference of different underbarrier paths. 

\section{CONCLUSION}
In the process of a conventional Wigner tunneling an electron encounters a classically allowed region, where the discrete 
energy level coincides with its energy. In our case a potential barrier is a constant in the direction of tunneling. But 
along the tunneling path the certain regions are formed, where, in the classical language, the kinetic energy of a motion 
perpendicular to tunneling is negative. These regions play a role of potential wells, where a discrete energy level can 
coincide with the electron energy. Such phenomenon, which occurs at the certain magnetic field, is called Euclidean 
resonance and substantially depends on a shape of potential forces in the direction perpendicular to tunneling. Under the 
conditions of Euclidean resonance a long distance underbarrier motion is possible. Euclidean resonance (formation of the 
long range coherence) constitutes a phenomenon which can be considered as an opposite pole with respect to Anderson 
localization (destruction of the long range coherence).  

\acknowledgments
I am grateful to B. Shklovskii and A. Vainshtein for valuable discussions.

\end{document}